# SPEAKER VERIFICATION IN MISMATCH TRAINING AND TESTING CONDITIONS [1]

*Marcos Faúndez-Zanuy(\*), Adam Slupinski (\*\*)*

(\*) Escola Universitària Politècnica de Mataró, SPAIN
(\*\*) University of Saarland, GERMANY
e-mail: faundez@eupmt.es http://www.eupmt.es

## ABSTRACT

This paper presents an exhaustive study about the robustness of several parameterizations, with a new database specially acquired for the purpose of a speaker recognition application. This database includes the following variations: different recording sessions (including telephonic and microphonic recordings), recording rooms, and languages (it has been obtained from a bilingual set of speakers). This study has been performed with covariance matrices in a text independent speaker verification application. It reveals that the combination of several parameterizations can improve the robustness in all the scenarios.

## 1. INTRODUCTION

Speech variability is a main degradation factor in speaker recognition tasks. For this reason, it is important to test speaker recognition algorithms in a wide range of situations, such as the ones we can find in a more realistic situation than the laboratory conditions. For this purpose, we have acquired a new database that allows the evaluation of a high number of variability factors.

Some papers have established the relevance of the recording conditions (noise, different room, microphone, etc.) in speaker recognition tasks, but it is not well known the relevance of the language in speaker recognition. The contribution of the language should be evaluated in comparative studies of different speaker verification methods: could the language yield to additional increments in recognition rates for any given method?. We have done a set of experiments with our new database in order to categorize this contribution.

For bilingual speakers in conversational speech is quite common the change from one language to the other, therefore it is interesting to evaluate if this fact can affect a speaker recognizer (phonetically there are significative differences between both languages. Mainly, the Catalan language has eight vowels and Spanish only five).

Using our database we have tested several parameterizations, and the main conclusion is that a robust system can be achieved by combining different methods.

Our previous work [1] reports similar results in a speaker identification application.

The paper is organized as follows: section 2 explains in detail the main characteristics of the database, and the sources of variability that can be modeled. Section 3 summarizes the robust parameterizations that have been evaluated. Section 4 deals with the verification algorithm. Section 5 summarizes the main results, and the improvements achieved using combinations between methods, and finally section 6 sets up the final conclusions.

## 2. NEW BILINGUAL DATABASE

The design of the speech corpus and its phonological and syllabic balance follows the parameters proposed in a first database [2]. The main relevant characteristics of our new recordings are the following:

1. 48 bilingual speakers (24 males & 24 females). The speech signal has been acquired at a sampling rate of 16KHz, and all the database is about 20 CD.

2. Four different recording sessions: S1 (first session), S2 (second session, recorded one week later), S3 (recorded another week later) and S4 (1 month after session S3). All the tasks have been sequentially collected in two languages (Catalan and Spanish) uttered from the same speaker. This has been done in all the session, so there are S1s, S1c, S2s. S2c, S3s, S3c, S4s, S4c (s=Spanish, c=Catalan).

3. Several tasks have been recorded in each session, including digits, sentences, text, etc. (as described in [2]). These tasks include specific text, different for each speaker, and common text for all the speakers.

4. Each task has been simultaneously recorded with two different microphones: AKG C420, AKG D40S for sessions 1 & 2, AKG C420, SONY ECM-66B for sessions 3 & 4, using one stereo channel for each microphone. In this paper we will use the following notation:

| M1 | AKG C420 |
|----|----------|
| M2 | AKG D40S |
| M3 | SONY ECM 66B |

**Table 1:** Microphones

---

[1] This work has been supported by the CICYT TIC97-1001-C02-02

5. Different recording conditions for all the tasks: One recording session in an anechoic room (S1AR=session one anechoic room), one recording session with a telephone handset plug into a PC connected to an ISDN (S3ISDN Session 3 ISDN recording).

## 3. PARAMETERIZATION METHODS

The speech material has been downsampled to 8KHz (except the S3ISDN that was originally captured at 8KHz A-law). Pre-emphasis of 0.95 was applied. Frames of 240 samples have been chosen (Hamming window), and an overlap between adjacent frames of 2/3. Frames under an energy threshold have been discarded. LPC coefficients have been obtained from each frame, using the Levinson-Durbin recursion. From these coefficients a recursion has been applied in order to obtain the LPCC (cepstral coefficients derived from the LPC coefficients).

The studied parameterizations are:

1. Cepstrum (LPCC) [3]
2. Cepstral mean subtraction [3] (CMS)

$$CMS = LPCC - \frac{1}{N}\sum_{i=1}^{N} LPCC_i$$, where $N$ is the number of frames.

3. Adaptive component weighted Cepstrum [3] (ACW-LPCC). This method is adaptive on a frame-by-frame basis.
4. Cepstral linear weighting [3] (LW-LPCC)

$$w(n) = \begin{cases} n & n=1,2,...,L \\ 0 & otherwise \end{cases}$$

5. Bandpass liftered cepstrum [3] (BPL-LPCC)

$$w(n) = \begin{cases} 1+\frac{1}{2}sin(\frac{n\pi}{L}) & n=1,2,...,L \\ 0 & otherwise \end{cases}$$

6. Cepstral std weighting (σ-LPCC) [4]. This parameterization consists on the normalization of each component of the cepstral vector, in order to achieve σ=1. This is an adaptive kind of cepstral weighting.
7. Postfilter Cepstrum [3] (PF-LPCC)

$$PF(n) = LPCC(n)(\alpha^n - \beta^n)$$, with α=1, β=0.9, for $n=1,...,P$

Some combinations between several of these parameterizations have also been tested. The tables of section 5 present the most successful ones. For instance, CMS+ACW is equivalent to the LPCC parameterization for each frame, and the sequential computation of Cepstral Mean Subtraction and Adaptive Component Weighting.

## 4. VERIFICATION ALGORITHM

Each speaker has been modeled with a Covariance matrix (CM). This model has been obtained with 1 minute of speech. In order to compute the distance of the input sentence to the model of the speaker whose identity is provided, an Arithmetic-harmonic sphericity measure [5] is applied:

$$d(C_j, C_{test}) = \log\left(tr(C_{test}C_j^{-1})tr(C_j C_{test}^{-1})\right) - 2\log(P)$$

where $tr$ is the trace of the matrix. The distance measure is converted into a likelihood by exponenting the utterance match scores: $L(C_j, C_{test}) = e^{-ad(C_j, C_{test})}$, where $a$ is a positive constant (we have chosen $a=2$). Likelihood ratios can then be formed using global speaker models or cohorts to normalize L.

The number of parameters for each speaker is $\frac{P^2 + P}{2}$ (the covariance matrix is symmetric). The speaker is accepted if $d(C_j, C_{test}) \leq T(j)$, where T(j) is the threshold for speaker j.

Using a set of five sentences (about 3 seconds per sentence) for each speaker, several thresholds are evaluated (100 values inside the range [0,1]), and the value that yields FAR=FRR is chosen. The resulting error is denominated Equal Error Rate (EER). It is possible that several points fulfill this condition, so several possibilities exist. Figure 1 shows an example of such situation.

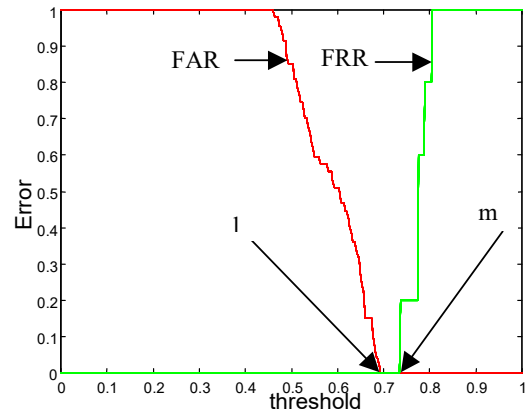

**Figure 1:** Example of possible thresholds for EER.

Although all the thresholds in the range [l,m] are valid, we have experimentally found that in this situation the most suitable value is l. Specially when the threshold is set in one condition and there is a mismatch in training and testing conditions. If the threshold is set to m or (l+m)/2 the FRR is dramatically increased.

Thus, we have used the following algorithm:

*l=minimun_likelihood_value;*

*while far(l)>frr(l), l=l+1; end*

In order to improve the results the cohort normalization technique [6] has been applied. Figure 2 shows the EER as function of the number of cohorts for S4c M3M3. On the other hand the greater the number of cohorts the greater the computational burden. In our simulations we have used 5 cohorts, chosen using the first test sentence of each speaker. From the set of cohorts, the following normalization is done:

$$L_j = L_j - max\{L_i, i \in cohorts\}$$

In a real situation the verification system is evaluated with testing sentences not used for computing the thresholds, and the arithmetic mean of the FAR and FRR is considered, rather than EER.

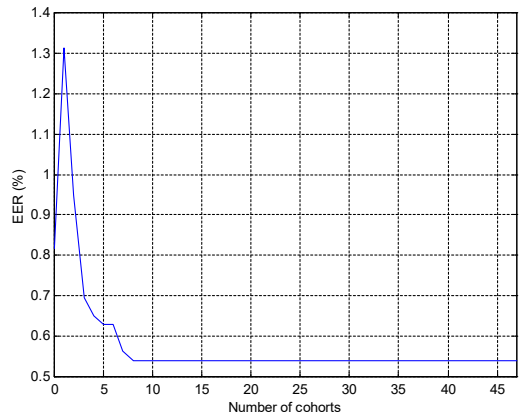

**Figure 2:** EER as function of the namber of cohorts por S4c M3M3

# 5. RESULTS

Our new bilingual database is suitable for the study of three relevant mismatch training and testing conditions: different microphones, temporal interval between training and recording sessions, and different languages. This section presents the most interesting results of our experiments, with LPCC vectors of dimension $P$=20 (matrices of size 20x20).

LPCC$_{3..P}$ means that the first and second coefficients have been removed.

## 5.1 Mismatch between training and testing microphones

In this section, the recording session and language are fixed (session 4 Catalan). We evaluate all the possible microphone combinations. For instance, M1M3 means training with microphone M1 and testing with microphone M3. Table 2 summarizes the identification rates using different microphones for S4c (session 4 Catalan), as function of the training and testing microphones. It is interesting to observe that although both microphones capture simultaneously the same acoustical signal, there are significative differences in M1M3 and M3M1 situations with respect to M1M1 and M3M3.

| PARAMETERIZ. | M1M1 | | M1M3 | | M3M3 | | M3M1 | |
|---|---|---|---|---|---|---|---|---|
| LPCC | 0.20 | 1.06 | 10.97 | 10.25 | 0.63 | 0.82 | 22.49 | 18.33 |
| LPCC$_{3..P}$ | 1.50 | 2.32 | 10.77 | 11.57 | 1.02 | 0.38 | 21.34 | 16.71 |
| σ-LPCC | 1.17 | 1.68 | 15.45 | 17.45 | 0.95 | 0.99 | 27.43 | 24.41 |
| ACW | 0.56 | 1.25 | 9.06 | 8.80 | 0.59 | 0.56 | 20.53 | 15.94 |
| CMS | 1.08 | 1.47 | 4.20 | 3.30 | 0.35 | 0.97 | 9.61 | 6.80 |
| CMS+ACW | 0.85 | 1.40 | 3.28 | 3.10 | 0.27 | 0.90 | 6.68 | 5.11 |
| CMS+ACW+σ-LPCC | 0.90 | 1.31 | 5.28 | 3.85 | 0.45 | 0.90 | 7.97 | 5.66 |
| CMS+σ-LPCC | 0.90 | 1.70 | 5.70 | 3.48 | 0.73 | 0.97 | 10.56 | 7.50 |
| CMS-LW | 1.08 | 1.47 | 4.20 | 3.30 | 0.35 | 0.97 | 9.61 | 6.80 |
| ACW+σ-LPCC | 1.08 | 1.47 | 15.40 | 16.61 | 0.90 | 1.02 | 26.64 | 23.46 |
| PF | 0.20 | 1.06 | 10.57 | 9.66 | 0.67 | 0.84 | 22.51 | 18.12 |
| CMS+PF | 1.08 | 1.47 | 3.76 | 3.32 | 0.56 | 0.97 | 9.61 | 6.83 |
| CMS+PF+σ-LPCC | 0.90 | 1.70 | 5.59 | 3.91 | 0.70 | 0.97 | 10.56 | 7.50 |

**Table 2:** EER (%) (with cohorts = 5 / without) for session 4 Catalan.

## 5.2 Mismatch between training and testing recording sessions.

In this section we have evaluated the relevance of different training and testing recording sessions, and the simultaneous mismatch of microphone and recording session.

Table 3 shows the identification rates in several conditions. For instance, S4cM1S2cM2 means that session 4 Catalan and microphone 1 are used for training, and Session 2 Catalan and microphone 2 for testing.

Obviously in real applications the training and testing sessions are not in the same day, so the identification rates can be degraded. The use of a robust parameterization can improve the results up to 9.2% (same microphone for testing and training), and up to 21% if there is also a change of microphone. Another approach consists on using several recording sessions for training the model of each speaker, instead of using only one recording session.

| PARAMETERIZ. | S4cM1 S3cM1 | | S4cM1 S2cM1 | | S4cM1 S2cM2 | | S2cM1 S4cM3 | |
|---|---|---|---|---|---|---|---|---|
| LPCC | 3.71 | 2.69 | 5.30 | 3.00 | 8.21 | 8.05 | 12.25 | 11.02 |
| LPCC$_{3..P}$ | 3.59 | 3.88 | 7.44 | 6.06 | 9.98 | 10.82 | 12.63 | 11.79 |
| σ-LPCC | 4.36 | 2.62 | 6.20 | 3.67 | 9.65 | 9.13 | 17.75 | 16.19 |
| ACW | 2.69 | 2.67 | 5.23 | 2.77 | 7.78 | 7.15 | 12.16 | 10.61 |
| CMS | 2.98 | 3.53 | 5.11 | 3.13 | 6.97 | 6.17 | 5.54 | 5.27 |
| CMS+ACW | 2.15 | 3.46 | 5.00 | 3.41 | 5.70 | 5.40 | 4.88 | 5.46 |
| CMS+ACW+σ-LPCC | 3.40 | 3.23 | 4.77 | 3.43 | 6.50 | 6.65 | 6.15 | 4.85 |
| CMS+σ-LPCC | 3.38 | 3.28 | 5.46 | 2.78 | 6.70 | 7.58 | 6.83 | 4.81 |
| CMS-LW | 2.98 | 3.53 | 5.11 | 3.13 | 6.97 | 6.17 | 5.54 | 5.27 |
| ACW+σ-LPCC | 3.67 | 2.85 | 6.88 | 3.75 | 9.64 | 8.79 | 17.52 | 16.27 |
| PF | 3.71 | 2.69 | 5.30 | 3.00 | 8.21 | 8.05 | 11.84 | 11.59 |
| CMS+PF | 2.98 | 3.53 | 5.11 | 3.13 | 6.97 | 6.17 | 5.35 | 5.23 |
| CMS+PF+σ-LPCC | 3.38 | 3.28 | 5.46 | 2.78 | 6.70 | 7.58 | 6.78 | 4.56 |

**Table 3:** EER (%) (with cohorts = 5 / without) for sessions 2,3 & 4 Catalan, (different training and testing sessions).

### 5.3 Mismatch training and testing recording languages.

This section presents the most relevant results in the following mismatch conditions: a) Training and testing languages, b) training and testing recording languages and microphones, c) Training and testing languages, microphones, and recording sessions.

Table 4 summarizes the verification error rates using different languages for training and testing (Catalan=c, Spanish=s), different microphones, and different sessions.

From this table it can be deduced that the change of recording language using the same microphone and recording session has a minor effect over the identification rates. This is because in a speaker identification system the goal is to model the characteristics of the speech production system, not the content of the message. In a similar way, persons can identify familiar voices although they are speaking in an unknown language. On the other hand, one language obtains better results than the other does, but there are small differences.

| PARAMETERIZ. | S4sM1 S4cM1 | | S4cM1 S4sM1 | | S4cM1 S4sM3 | | S2sM1 S4cM3 | |
|---|---|---|---|---|---|---|---|---|
| LPCC | 0.40 | 1.83 | 1.76 | 2.80 | 10.40 | 11.15 | 13.17 | 13.68 |
| LPCC$_{3..P}$ | 1.50 | 2.32 | 2.90 | 3.30 | 11.27 | 9.06 | 10.04 | 11.09 |
| $\sigma^2$-LPCC | 1.28 | 2.1 | 2.37 | 3.54 | 18.06 | 20.96 | 18.28 | 22.52 |
| ACW | 0.27 | 1.26 | 1.21 | 2.24 | 10.04 | 9.78 | 12.04 | 11.8 |
| CMS | 0.98 | 2.19 | 1.67 | 4.02 | 5.25 | 7.15 | 5.87 | 9.16 |
| CMS+ACW | 0.81 | 1.92 | 1.58 | 3.59 | 4.04 | 6.11 | 5.14 | 8.65 |
| CMS+ACW+$\sigma$-LPCC | 4.34 | 3.91 | 1.47 | 4.05 | 4.24 | 6.54 | 5.64 | 8.48 |
| CMS+$\sigma$-LPCC | 0.98 | 2.05 | 2.31 | 4.04 | 5.88 | 7.75 | 6.99 | 8.71 |
| CMS-LW | 0.98 | 2.19 | 1.67 | 4.02 | 5.25 | 7.15 | 5.87 | 9.02 |
| ACW+$\sigma$-LPCC | 1.10 | 1.94 | 1.96 | 3.24 | 17.13 | 19.30 | 17.15 | 21.64 |
| PF | 0.40 | 1.83 | 1.76 | 3.01 | 10.70 | 11.30 | 12.53 | 13.67 |
| CMS+PF | 0.98 | 2.19 | 1.67 | 3.82 | 5.10 | 6.92 | 5.68 | 9.41 |
| CMS+PF+$\sigma$-LPCC | 0.98 | 2.05 | 2.31 | 4.04 | 5.83 | 7.54 | 6.43 | 9.23 |

**Table 4:** EER (%) (with cohorts = 5 / without) for different training and testing languages.

Finally, table 5 shows the value of ½(FAR+FRR) when the thresholds are fixed with S4cM1M1. The results have been obtained with the most robust parameterization ( CMS + ACW) against mismatch in speaker identification task [1].

From table 5 it can be deduced that if the threshold is set up without mismatch, the verification errors increase significantly.

| | S4cM1 S4cM3 | | S4cM1 S2cM1 | | S4cM1 S2cM2 | | S4cM1 S4sM1 | | S4cM1 S4sM3 | |
|---|---|---|---|---|---|---|---|---|---|---|
| FAR | 1.99 | 1.29 | 1.73 | 2.75 | 1.95 | 3.46 | 1.37 | 8.24 | 1.99 | 3.68 |
| FRR | 14.2 | 48.8 | 30.83 | 39.17 | 13.75 | 16.25 | 4.58 | 6.25 | 20.83 | 50 |
| ½(FAR+FRR) | 8.08 | 25.0 | 16.28 | 20.96 | 7.85 | 9.85 | 2.98 | 7.25 | 11.41 | 26.84 |

**Table 5:** Threshold fixed with S4cM1M1 (with cohorts = 5 / without)

Table 6 shows the results in similar conditions than table 1, but using S4cM1S2cM2 for setting up the thresholds. It can be seen that it is better to fix the threshold with a mismatch test session (table 6) for improving the results in mismatch conditions (see the difference between tables 5 and 6.

| | S4cM1 S4cM1 | | S4cM1 S4cM3 | | S4cM1 S2cM1 | | S4cM1 S4sM1 | | S4cM1 S4sM3 | |
|---|---|---|---|---|---|---|---|---|---|---|
| FAR | 3.95 | 4.39 | 5.67 | 1.68 | 4.88 | 3.95 | 4.17 | 11.6 | 5.63 | 4.43 |
| FRR | 2.08 | 5.42 | 8.75 | 47.5 | 20 | 29.58 | 1.67 | 4.17 | 11.67 | 44.58 |
| ½(FAR+FRR) | 3.01 | 4.90 | 7.21 | 24.59 | 12.44 | 16.76 | 2.92 | 7.89 | 8.65 | 24.51 |

**Table 6:** Threshold fixed with S4cM1S2cM2 (with cohorts = 5 / without)

## 6. CONCLUSIONS

In this paper we have studied the relevance of several mismatch conditions in a speaker verification application. It is important because in real applications it is quite difficult to obtain the same conditions in train and test phases (different kinds of microphones, test sentences are recorded in different days than the training sentences, for bilingual speakers it is quite common to change from one language to the other, etc.). Our study includes an exhaustive study of several parameterizations and combinations between them in order to obtain the most robust features to mismatch conditions in all the scenarios. The results of section 5 let us to establish the following conclusions:

- The change of microphone between training and testing sessions is more important than the mismatch of recording session and the mismatch of recording languages.
- The parameterization algorithm (implemented preprocessing algorithm over the LPCC coefficients) is more relevant than the change of the identification algorithm. In fact, the reported differences in identification rates against the change of identification algorithm that can be found in the literature are smaller than the change of the parameterization.
- Although it is possible to find an optimal parameterization for each particular condition, the best parameterization in all the scenarios seems to be the CMS+ACW.